\documentclass[iop]{emulateapj}
\usepackage{natbib}
\usepackage{graphicx}
\usepackage{times}
\usepackage{xspace} 
\usepackage{booktabs} 

\slugcomment{\it Accepted version -- May 2015}
\shorttitle{X-Ray to Mid-IR Relation at High Luminosity}
\shortauthors{Stern}

\def\ie{{i.e.}}
\def\eg{{e.g.}}

\def\wise{{\it WISE}}

\def\chandra{{\it Chandra}}
\def\xmm{{\it XMM-Newton}}
\def\spitzer{{\it Spitzer}}
\def\nustar{{\it NuSTAR}}

\def\deg{\ifmmode {^{\circ}}\else {$^\circ$}\fi}
\def\kms{\ifmmode {\rm\,km\,s^{-1}}\else
    ${\rm\,km\,s^{-1}}$\fi}
\def\ergcm2s{\ifmmode {\rm\,erg\,cm^{-2}\,s^{-1}}\else
    ${\rm\,erg\,cm^{-2}\,s^{-1}}$\fi}
\def\ergAcm2s{\ifmmode {\rm\,erg\,cm^{-2}\,s^{-1}\,\AA^{-1}}\else
    ${\rm\,erg\,cm^{-2}\,s^{-1}\,\AA^{-1}}$\fi}
\def\ergs{\ifmmode {\rm\,erg\,s^{-1}}\else
    ${\rm\,erg\,s^{-1}}$\fi}
\def\kmsMpc{\ifmmode {\rm\,km\,s^{-1}\,Mpc^{-1}}\else
    ${\rm\,km\,s^{-1}\,Mpc^{-1}}$\fi}

\def\spose#1{\hbox to 0pt{#1\hss}}
\def\simlt{\mathrel{\spose{\lower 3pt\hbox{$\mathchar"218$}}
     \raise 2.0pt\hbox{$\mathchar"13C$}}}
\def\simgt{\mathrel{\spose{\lower 3pt\hbox{$\mathchar"218$}}
     \raise 2.0pt\hbox{$\mathchar"13E$}}}

\def\plotfiddle#1#2#3#4#5#6#7{\centering \leavevmode
\vbox to#2{\rule{0pt}{#2}}
\includegraphics{#1}}

\begin{document}

\title{The X-Ray to Mid-Infrared Relation of AGN at High Luminosity}

\author{Daniel~Stern\altaffilmark{1}}
 
\altaffiltext{1}{Jet Propulsion Laboratory, California Institute
of Technology, 4800 Oak Grove Drive, Mail Stop 169-221, Pasadena,
CA 91109, USA [e-mail: {\tt daniel.k.stern@jpl.nasa.gov}]}

\begin{abstract} 

The X-ray and mid-IR emission from active galactic nuclei (AGN) are
strongly correlated.  However, while various published parameterizations
of this correlation are consistent with the low-redshift, local
Seyfert galaxy population, extrapolations of these relations to
high luminosity differ by an order of magnitude at $\nu L_\nu$(6
$\mu$m) $\sim 10^{47}\, {\rm erg}\, {\rm s}^{-1}$.  Using data from
the {\it Wide-field Infrared Survey Explorer}, we determine the
mid-IR luminosities of the most luminous quasars from the Sloan
Digital Sky Survey and present a revised formulation of the
X-ray to mid-IR relation of AGN which is appropriate from the Seyfert
regime to the powerful quasar regime.

\end{abstract}


\section{Introduction}

Several groups have investigated how the X-ray and mid-IR emission
from active galactic nuclei (AGN) are correlated.  \citet{Lutz:04}
and \citet{Gandhi:09} investigated local samples of Seyfert galaxies,
establishing this correlation at low luminosities.  \citet{Fiore:09}
investigated unobscured (or type-1) AGN at higher luminosities
identified in the COSMOS and {\it Chandra} Deep Field-South (CDF-S)
fields, while \citet{Lanzuisi:09} investigated obscured (or type-2)
AGN at similar luminosities identified in \spitzer\ fields.  However,
while these various efforts have presented X-ray to mid-IR correlations
that are largely consistent at low luminosity, extrapolation of
these results to high luminosity differ by an order of magnitude
at $\nu L_\nu$(6 $\mu$m) $\sim 10^{47}\ {\rm erg}\ {\rm s}^{-1}$.

This can be particularly problematic for investigations into obscured,
high-luminosity AGN.  Obscuration will preferentially affect UV and
(low-energy) X-ray emission, while the mid-IR emission is largely
unaffected until the most extreme obscuring columns are attained
($A_V \sim 30$).  Several investigations have therefore adopted the
mid-IR luminosity of AGN as a robust indicator of the intrinsic AGN
strength, and then used the relative luminosity at X-ray energies
to measure the amount of obscuration \citep[\eg,][]{Fiore:08,
Fiore:09, Georgantopoulos:11, Luo:13, Lansbury:14, Rovilos:14,
Stern:14}.  However, with substantial discrepancies between published
intrinsic X-ray to mid-IR luminosity ratios for luminous quasars,
this leads to uncertainties in such analyses.

Using data from the {\it Wide-field Infrared Survey Explorer}
\citep[\wise;][]{Wright:10}, we determine the mid-IR luminosities
of a sample of extremely luminous unobscured quasars from
\citet{Just:07}, who reported on their X-ray properties.  The sample
consists of essentially all the most optically luminous quasars
known at the time ($M_i \approx -29.3$ to $-30.2$).  We supplement
these data with published X-ray and mid-IR luminosities of sources
at lower luminosities, including published high-resolution mid-IR
imaging of nearby Seyfert galaxies observed with VLT which allow
for robust separation of the nuclear emission from host galaxy
emission.   Combining these various data sets, we present a revised
X-ray to mid-IR relation of AGN spanning from low luminosity Seyfert
galaxies to the most powerful quasars known.  Throughout, we adopt
the concordance cosmology, $\Omega_{\rm M} = 0.3$, $\Omega_\Lambda
= 0.7$ and $H_0 = 70\, \kmsMpc$.

\section{Samples and Mid-IR Luminosities}

This section describes several AGN samples with mid-IR and X-ray
data publicly available, which we use to investigate the relation
between AGN luminosities in these two energy ranges.  A concern in
such work is host galaxy contamination to the AGN luminosity.  This
is less problematic at higher energies as 0.5-10~keV X-ray luminosities
above $3 \times 10^{42}\, {\rm erg}\, {\rm s}^{-1}$ are predominantly
due to AGN, with only rare, extreme starbursts in the distant
universe as exceptions \citep[see review by][]{Brandt:15}.  At
mid-IR wavelengths, host galaxy contamination can be a larger
concern, particularly for lower luminosity samples as the AGN becomes
a minor contributor to the overall IR luminosity of a galaxy.
For these lower luminosity sources, we rely on published work which
separates the host and nuclear emission either through high-resolution
imaging \citep{Gandhi:09} or spectral decomposition \citep{Lutz:04}.
At higher luminosities, entering into the quasar regime, host galaxy
contamination becomes less of a concern.  We adopt a traditional
line at 2-10~keV X-ray luminosities of $10^{43}\, {\rm erg}\, {\rm
s}^{-1}$ to identify quasars, with the caveat that mid-IR spectra
of sources at these luminosities do sometimes show star-formation
features, implying that a measurable fraction of the mid-IR emission
is not related to the central engine.

Regarding that latter point, \citet{Murphy:09} and \citet{Fadda:10}
present deep \spitzer\ mid-IR spectroscopy of IR-selected
galaxies from the GOODS-N and GOODS-S fields, respectively.  
\citet{Fadda:10} finds that galaxies with 2-8~keV luminosities
$\simgt 10^{44}\, {\rm erg}\, {\rm s}^{-1}$ are strongly AGN dominated
in the mid-IR.  \citet{Murphy:09} includes three galaxies with
2-8~keV luminosities between $10^{43}\, {\rm erg}\, {\rm s}^{-1}$
and $5 \times 10^{43}\, {\rm erg}\, {\rm s}^{-1}$, all three of
which show star formation contributions to the mid-IR luminosity
as evidenced by polycyclic aromatic hydrocarbon (PAH) features.
Based on mid-IR spectral modeling, \citet{Murphy:09} reports mid-IR
AGN fractions of $\sim 50\%$ for all three galaxies.  While this
is somewhat a concern, we note that the majority of the high
luminosity sources in our analysis detailed below are considerably
more luminous, where such contributions should be less.  Finally,
we emphasize that the above work on mid-IR spectra were from
IR-selected samples, not X-ray or optically selected samples, so
would be biased towards sources where mid-IR emission from star
formation was strongest.  


\subsection{High Luminosity Sample}

\citet{Just:07} reports on the X-ray properties of the most luminous
quasars from the Sloan Digital Sky Survey (SDSS) using a mixture
of dedicated {\it Chandra} observations and archival data from {\it
Chandra}, {\it XMM-Newton} and {\it ROSAT}.  The core sample are
the 32 quasars in the SDSS Data Release 3 (DR3) quasar catalog
\citep{Schneider:05} with $M_i < -29.28$.  The quasars are in the
redshift range $1.5 < z < 4.6$, and the sample is supplemented
with two luminous quasars in the DR3 survey area that were
missed by SDSS. 

In our analysis, we exclude several of the sources from the core
sample (Table~1).  First, we exclude known gravitationally lensed
quasars, for which the measured luminosities are not representative
of the intrinsic values.  We exclude broad absorption line (BAL)
quasars, which are known to often exhibit suppressed X-ray emission
\citep[\eg,][]{Gallagher:02, Luo:13, Luo:14, Teng:14}.  Following
\citet{Just:07}, we exclude the anomalously X-ray weak quasar
SDSS~J1521+5202 which lacks Ly$\alpha$ emission; SDSS~J1521+5202
instead exhibits strong Ly$\alpha$ absorption, suggestive of absorbing
material along the line-of-sight.  We also exclude SDSS~J1350+5716
and SDSS~J1421+4633, which only have two and four \chandra\ counts,
respectively.  The remaining \chandra-observed sources in our sample
have between 10 and 348 counts in the typically 4~ks \chandra\
observations, with a median value of 41 counts.  Only two sources
have less than 20 counts (SDSS~J1438+4314 -- 10 counts; SDSS~J0209--0005
-- 18 counts).

Table~1 presents the rest-frame 2-10~keV X-ray luminosities of the
\citet{Just:07} sample.  The values, which come directly from that
work, have been corrected for the Galactic absorption to each source,
as well as the quantum efficiency decay of \chandra\ at low energies.
The fluxes were calculated using {\tt PIMMS}, assuming a power-law
model with $\Gamma = 2.0$, which is a typical photon index for
luminous AGN \citep[\eg,][]{Reeves:00, Vignali:05}.  This assumes
the intrinsic X-ray absorption to be negligible for these luminous,
broad-lined quasars.  This assumption is supported by the fact that
the observed band ratios presented in \citet{Just:07} are consistent
with $\Gamma = 2.0$ for most of the culled sample (\eg, $\Gamma =
2.0$ is within the $1 \sigma$ range for 16 of the 24 sources and
within the $2 \sigma$ range for 22 sources).  The X-ray spectral
fitting finds $\Gamma$ significantly different from 2.0 for only
two of the included sources, SDSS~J1614+4704 ($\Gamma = 1.5 \pm
0.2$) and SDSS~J2123--0050 ($\Gamma > 2.1$), but our results are
essentially unchanged whether or not we include these two sources.

\begin{figure}
\plotfiddle{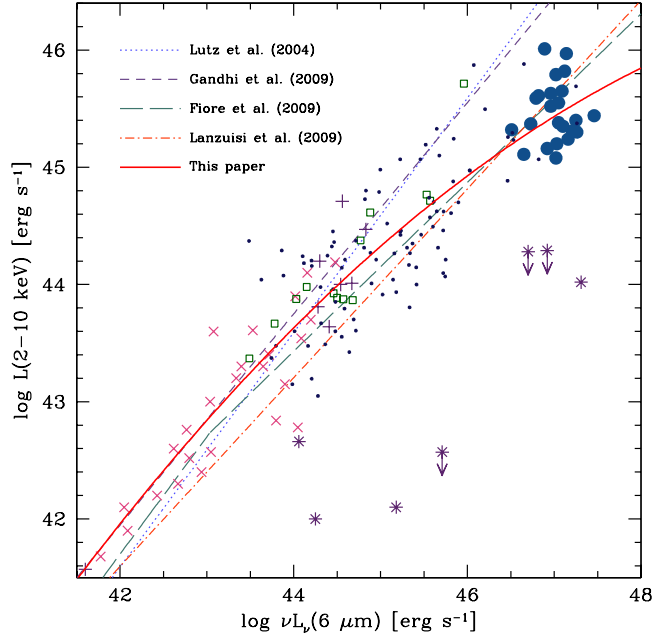}{3.4in}{0}{45}{45}{-133}{-70}
\caption{Rest-frame 2-10~keV X-ray luminosity against rest-frame
6$\mu$m luminosity for a sample of AGN with four published relations
plotted (as indicated), as well as our new derivation of the X-ray
to mid-IR relation (solid red).  Large solid blue circles show
luminous quasars from \citet{Just:07}.  Open green squares show
broad-lined AGN from the SEXSI survey \citet{Eckart:10}.  Purple
plus signs show Compton-thin AGN identified by \nustar\
\citep{Alexander:13, DelMoro:14}, while purple asterisks show
candidate Compton-thick AGN studied by \nustar\ \citep[][see text
for details]{Gandhi:14, Lansbury:14, Stern:14}.  Small blue dots
show quasars from SDSS DR5 with X-ray data reported in \citet{Young:09}.
Red exes show local Seyfert galaxies from \citet{Horst:08} and
\citet{Gandhi:09}.
\label{fig:fig1}}
\end{figure}

\begin{figure}
\plotfiddle{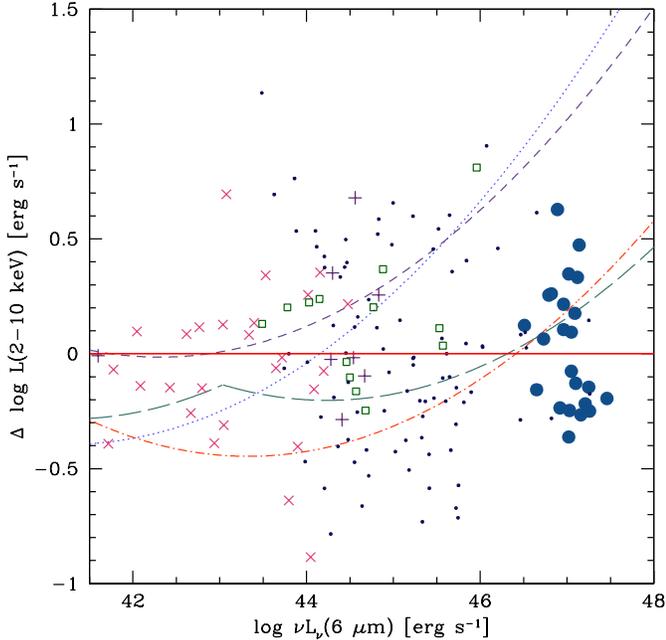}{3.4in}{0}{45}{45}{-133}{-70}
\caption{Difference between rest-frame 2-10~keV X-ray luminosity
and the value predicted by the relation presented herein, plotted
against rest-frame 6$\mu$m luminosity.  Symbols and published
relations are plotted as per Figure~1.
\label{fig:fig2}}
\end{figure}

We determine the rest-frame $6 \mu$m luminosities for this sample
of luminous quasars using \wise.  We exclude SDSS~J1733+5400 for
which the mid-IR photometry is corrupted from the scattered light
halo of a nearby bright source ({\tt ccflag} = hHHH).  The remaining
23 core sample quasars have robust mid-IR photometry from \wise.
For the $12 \mu$m channel of \wise\ ($W3$), the signal-to-noise
ratio of the core sample ranges from 9.9 to 58.4, with a median
value of $\langle SNR_{\rm W3} \rangle = 29.8$.  For the $22 \mu$m
channel of \wise\ ($W4$), the corresponding range is 5.0 to 26.4,
with a median value of $\langle SNR_{\rm W4} \rangle = 10.5$.   For
redshifts $z < 3.7$, which includes all but the three highest
redshift quasars in the core sample, rest-frame $6 \mu$m is within
the $W3$ to $W4$ wavelength range.  We therefore simply linearly
interpolate (or extrapolate for the $z > 3.7$ quasars) the $W3$ and
$W4$ photometry to determine the rest-frame $6 \mu$m luminosity,
$\nu L_\nu (6 \mu{\rm m})$.  Table~1 presents luminous quasar sample
with their redshifts, mid-IR luminosities and absorption-corrected,
rest-frame 2-10~keV X-ray luminosities from \citet{Just:07}.
Figure~\ref{fig:fig1} plots these rest-frame 2-10~keV luminosities
against rest-frame $6 \mu$m luminosity, and Figure~\ref{fig:fig2}
plots the residuals of these points relative to the new X-ray to
mid-IR relation derived in \S~3.2.

In their analysis of the X-ray to optical properties of luminous
quasars, \citet{Just:07} also include a complementary sample of $z
\simgt 4$ quasars with $M_i \simlt -29$ from outside the SDSS DR3
quasar region with available X-ray data.  The vast majority of these
sources are at such great distance that they are faint or undetected
by \wise\ (\eg, $SNR_{\rm W4} < 3$).  We therefore do not include
this sample in our analysis.

\scriptsize
\begin{deluxetable}{lcccl}
\tablecaption{High luminosity sample.}
\tablehead{
\colhead{SDSS ID} &
\colhead{$z$} &
\colhead{$\log \nu L_\nu (6 \mu{\rm m})$} &
\colhead{$\log L_{2-10}$} &
\colhead{Notes}}
\startdata
J012156.04+144823.9 & 2.87 & 46.96 & 45.52 & \\ 
J014516.59--094517.3 & 2.73 & 46.87 & 46.13 & lensed \\
J020950.71--000506.4 & 2.85 & 47.16 & 45.24 & \\ 
J073502.31+265911.4 & 1.97 & 47.02 & 45.08 & \\ 
J075054.64+425219.2 & 1.90 & 46.92 & 45.16 & \\ 
J080342.04+302254.6 & 2.03 & 46.51 & 45.32 & \\ 
J081331.28+254503.0 & 1.51 & 47.20 & 45.68 & lensed \\
J084401.95+050357.9 & 3.35 & 47.10 & 45.31 & BAL \\ 
J090033.49+421546.8 & 3.29 & 47.14 & 45.97 & \\ 
J094202.04+042244.5 & 3.28 & 46.79 & 45.59 & \\ 
J095014.05+580136.5 & 3.96 & 47.05 & 45.55 & \\ 
J100129.64+545438.0 & 1.76 & 46.65 & 45.11 & \\ 
J101447.18+430030.1 & 3.13 & 47.05 & 45.38 & \\ 
J110610.73+640009.6 & 2.20 & 47.09 & 45.65 & \\ 
J111038.64+483115.6 & 2.96 & 47.26 & 45.30 & \\ 
J121930.77+494052.3 & 2.70 & 47.02 & 45.79 & \\ 
J123549.47+591027.0 & 2.82 & 46.73 & 45.37 & \\
J123641.46+655442.0 & 3.39 & 47.10 & 45.35 & \\
J135044.67+571642.8 & 2.91 & 46.48 & 44.31 & excluded \\ 
J140747.22+645419.9 & 3.08 & 46.82 & 45.61 & \\ 
J142123.98+463317.8 & 3.37 & 47.10 & 44.66 & excluded \\ 
J142656.17+602550.8 & 3.19 & 47.46 & 45.44 & \\ 
J143835.95+431459.2 & 4.61 & 47.21 & 45.31 & \\
J144542.75+490248.9 & 3.88 & 46.89 & 46.01 & \\ 
J152156.48+520238.4 & 2.19 & 47.04 & 43.95 & excluded \\ 
J152553.89+513649.1 & 2.88 & 46.93 & 45.92 & BAL \\
J161434.67+470420.0 & 1.86 & 46.96 & 45.63 & \\ 
J162116.92--004250.8 & 3.70 & 47.12 & 45.82 & \\
J170100.62+641209.0 & 2.74 & 47.25 & 45.40 & \\
J173352.22+540030.5 & 3.43 & 46.52 & 45.56 & excluded \\ 
J212329.46--005052.9 & 2.26 & 47.03 & 45.20 & \\ 
J231324.45+003444.5 & 2.08 & 46.84 & 44.62 & BAL 
\enddata
\label{table:just07}
\tablecomments{Luminosities are in units of erg\, s$^{-1}$.  BAL
indicates broad absorption line quasars, which, along with known
gravitationally lensed quasars, are excluded in the analysis.  See
\S2.1 for details on the other excluded sources.  Luminosities are
in the rest-frame, and 2-10~keV X-ray luminosities, from \citet{Just:07},
are absorption-corrected.}
\end{deluxetable}
\normalsize

\subsection{SEXSI Sample}

The Serendipitous Extragalactic X-ray Source Identification (SEXSI)
program surveyed 2-10~keV selected sources from more than 2~deg$^2$
identified serendipitously in several dozen extragalactic pointings
by {\it Chandra} \citep{Harrison:03, Eckart:05, Eckart:06, Eckart:10}.
Out of a total sample of 1034 sources, SEXSI obtained nearly 500
spectroscopic identifications of sources with intermediate hard
X-ray fluxes, $S({\rm 2-10~keV)} \sim 10^{-13}$ to $10^{-15}\, {\rm
erg}\, {\rm cm}^{-2}\, {\rm s}^{-1}$. SEXSI complements the {\it
Chandra} Deep Fields, which reach depths more than an order magnitude
fainter, but cover a total area of less than 0.2~deg$^2$.  In
Figure~1 we plot the X-ray and mid-IR properties of a subset of
SEXSI sources from \citet{Eckart:10}, which reported on {\it Spitzer}
mid-IR follow-up of approximately one-third of the SEXSI survey
fields.  Specifically, \citet{Eckart:10} provides mid-IR photometry
for 290 of the 1034 SEXSI sources.  We only plot sources whose
optical spectra are classified as broad-lined AGN \citep[65 of the
290 sources in][]{Eckart:10}.  Since we are interested in rest-frame
$6 \mu$m luminosities, we require sources have good photometry at
both $8$ and $24 \mu$m ({\tt flag} = 1 in Eckart et al. 2010,
implying robust counterparts with $\geq 5\sigma$ detections,
unaffected by nearby bright sources), from which we interpolate the
mid-IR fluxes to determine luminosities at rest-frame $6 \mu$m.
This further reduces the sample to 25 targets.  In order to avoid
low-luminosity sources for which the mid-IR photometry might be
contaminated by host galaxy light, we require $L({\rm 2-10~keV})
\geq 10^{43}\, {\rm erg}\, {\rm s}^{-1}$, which eliminates two
sources.  We also require robust hard band detections, with $SNR_{\rm
2-10} \geq 3$, which eliminates seven additional sources, leaving
a sample of 16 sources.

The plotted X-ray luminosities are based on the Galactic
absorption-corrected 2-10~keV luminosities reported in \citet{Eckart:10},
adjusted to both account for the slightly different cosmology adopted
in that work as well as to correct for the intrinsic absorption
measured in that work based on X-ray spectral fitting with {\tt
XSPEC}.  We assume an intrinsic power-law X-ray spectrum with $\Gamma
= 1.9$, as typical of quasars at this luminosity, to derive this
final correction factor.  As a final clean-up of the sample, we
drop the three most heavily obscured sources based on the spectral
modeling, all sources with $N_{\rm H} \sim 10^{23}\, {\rm cm}^{-2}$,
corresponding to correction factors $> 65\%$.  The median neutral
column of the remaining 13 sources is $\langle N_{\rm H} \rangle =
2.5 \times 10^{21}\, {\rm cm}^{-2}$, and the median obscuration
correction factor is 2.4\%.  All remaining sources have correction
factors $< 15\%$, and 9 of the 13 have correction factors $< 5\%$.
In the end, this leaves a total sample of 13 broad-lined SEXSI
quasars with robust photometry at both X-ray and mid-IR energies
(see Table~\ref{table:sexsi}).  As expected, the broad-lined AGN
in SEXSI are at lower luminosities than the extreme sources from
\citet{Just:07}.

\scriptsize
\begin{deluxetable}{lccc}
\tablewidth{0pt}
\tablecaption{SEXSI sample.}
\tablehead{
\colhead{SEXSI ID} &
\colhead{$z$} &
\colhead{$\log \nu L_\nu (6 \mu{\rm m})$} &
\colhead{$\log L_{2-10}$}}
\startdata
J084854.4+445149 & 1.03 & 44.50 & 43.89 \\
J084858.0+445434 & 0.57 & 44.68 & 43.87 \\
J091027.0+542054 & 1.64 & 44.77 & 44.38 \\
J091028.9+541523 & 0.65 & 43.78 & 43.67 \\
J091041.4+541945 & 0.79 & 44.03 & 43.88 \\
J091059.4+541715 & 1.86 & 44.88 & 44.61 \\
J091100.2+542540 & 1.89 & 45.57 & 44.71 \\
J133810.9+293119 & 2.03 & 45.53 & 44.77 \\
J171614.4+671344 & 1.13 & 44.46 & 43.93 \\
J171635.5+671626 & 0.50 & 43.49 & 43.37 \\
J171638.0+671155 & 1.33 & 44.15 & 43.98 \\
J224716.9+033432 & 3.82 & 45.96 & 45.71 \\
J224731.6+033550 & 1.00 & 44.57 & 43.87 \\
\enddata
\label{table:sexsi}
\tablecomments{Luminosities are in the rest-frame, in units of erg\,
s$^{-1}$. X-ray luminosities are from Eckart et al. (2010), corrected
for intrinsic absorption reported in that work and adjusted for the
different cosmology adopted here.  See \S2.2 for details.}
\end{deluxetable}
\normalsize

\subsection{SDSS DR5 Sample}

\citet{Young:09} cross-matched the SDSS Data Release 5 (DR5) quasar
catalog with the {\it XMM-Newton} public archive.  A total of 792
quasars were matched, and \citet{Young:09} provide basic X-ray
spectral fits for each source.  For our analysis, we again exclude
BAL quasars, and we only consider sources with $\geq 6 \sigma$ X-ray
detections that were best fit with a single, non-absorbed power-law
model.  For all such sources, \citet{Young:09} also fit an absorbed
power-law model and provide the 90\% upper limit to the intrinsic
absorption, $N_{\rm H}$.  In order to avoid possibly obscured
quasars, we require that the value of that parameter be less than
$10^{21}\, {\rm cm}^{-2}$ such that intrinsic absorption is negligible
at rest-frame 2-10~keV.  Similar to the SEXSI sample, we also require
$L({\rm 2-10~keV}) \geq 10^{43}\, {\rm erg}\, {\rm s}^{-1}$ in order
to avoid low-luminosity systems which might have more host galaxy
contamination.    We then cross-matched this subset of 441 quasars
to the \wise\ archive, restricting our analysis to sources detected
with robust, $\geq 5 \sigma$ detections at both 12 $\mu$m (78\% of
the sample) and 22 $\mu$m (28\% of the sample), as well as clean
\wise\ photometry ({\tt ccflag} = 0000).  This requirement of robust
mid-IR detections might introduce concerns with a bias towards high
mid-IR emission quasars. However, the similar location of the SDSS
DR5 quasars in Figure~1 to the SEXSI quasars of comparable X-ray
luminosity but significantly deeper mid-IR data suggests this is
not the case.  Instead, the robust 22 $\mu$m detections essentially
biases the sample to lower redshift.  Whereas the initial subset
of 441 quasars has a median redshift $\langle z \rangle = 1.705$,
the final, conservative sample of 90 SDSS DR5 quasars with robust
X-ray and mid-IR detections, plotted in Figure~1 and listed in
Table~3, has a median redshift of $\langle z \rangle = 0.618$.  This
plotted sample bridges the high luminosity quasars discussed in \S
2.1 and the local, low-luminosity samples discussed next in \S 2.4.

As a back-of-the-envelope exercise, we use the large, mid-luminosity
SDSS quasar sample to investigate whether star formation contamination
to the mid-IR luminosities could be impacting our results.  {\it
Herschel} studies show that AGN, on average, have similar star
formation rates to massive `main sequence' galaxies at similar
redshift \citep[\eg,][]{Mullaney:12, Santini:12, Rosario:13}.  We
therefore use the specific star formation rate of main sequence
galaxies as a function of redshift determined by \citet{Elbaz:11},
and assume host galaxy masses of $\log(M / M_\odot) = 10.5$ in order
to derive typical star formation rates for the SDSS quasars.  We
then adopt the \citet{Rieke:09} conversion between star formation
rate and $24 \mu$m luminosity.  Based on the empirical star-forming
galaxy templates of \citet{Assef:10}, we shift these $24 \mu$m
luminosities downwards by a factor of 1.5 to determine the estimated
star-formation-related $6 \mu$m luminosities for the quasars in our
sample.  We then compare this to the observed $6 \mu$m luminosities.
We find that the $6 \mu$m luminosities are strongly dominated by
the AGN, with star formation accounting for just 1-2\% of the
observed $6 \mu$m luminosities, on average.  In only one quasar
does the inferred star formation contribute more than 50\% of the
observed $6 \mu$m luminosity, and in only three quasars does the
inferred star formation contribute more than 33\% of the observed
$6 \mu$m luminosity.  Assuming the above assumptions are valid,
\ie, that the SDSS DR5 subsample of quasars are hosted by massive,
main sequence galaxies, we therefore conclude that star formation
contamination will, on average, have a negligible contribution to
the $6 \mu$m luminosities of the quasars in our sample and not
affect the final X-ray to mid-IR relation derived below.

\scriptsize
\begin{deluxetable}{lccc}
\tablewidth{0pt}
\tablecaption{SDSS DR5 sample.}
\tablehead{
\colhead{SDSS ID} &
\colhead{$z$} &
\colhead{$\log \nu L_\nu (6 \mu{\rm m})$} &
\colhead{$\log L_{2-10}$}}
\startdata
J020011.52--093126.1 & 0.360 & 44.21 & 44.20 \\ 
J020118.67--091935.7 & 0.661 & 45.24 & 44.46 \\
J023402.08--084314.6 & 1.264 & 45.73 & 44.62 \\
J074020.22+311841.2  & 0.296 & 43.80 & 43.48 \\
J080608.13+244421.0  & 0.358 & 44.45 & 43.96 \\
J080711.01+390419.7  & 0.369 & 44.21 & 44.16 \\
J081422.12+514839.4  & 0.377 & 44.72 & 44.37 \\
J082257.55+404149.7  & 0.865 & 45.84 & 44.88 \\
J091029.02+542719.0  & 0.526 & 44.59 & 44.21 \\
J091301.03+525928.9  & 1.377 & 46.65 & 45.88 \\
\enddata
\label{table:sdss}
\tablecomments{Luminosities are in the rest-frame, in units of erg\,
s$^{-1}$.  X-ray luminosities are from \citet{Young:09}.  This table
is published in its entirety in the electronic edition of ApJ; a
portion is shown here for guidance regarding its form and content.}
\end{deluxetable}
\normalsize

\subsection{Low Luminosity Samples}

\citet{Horst:08} and \citet{Gandhi:09} report on near-diffraction-limited
$12 \mu$m imaging of local Seyfert galaxies obtained with the VLT
Imager and Spectrometer for the mid-IR \citep[VISIR;][]{Lagage:04}.
The high spatial resolution allows them to measure the mid-IR
luminosities of the galactic nuclei with minimal contamination from
star formation in the outer galaxy.  They also provide the intrinsic
2-10~keV luminosities for their sample with corrections applied
for non-nuclear components, obscuration, and the modest redshift
effects necessary for this local sample.  In our study, we exclude
the Compton-thick sources flagged in \citet{Gandhi:09} for which
the intrinsic X-ray luminosities are more model dependent.  We also
exclude LINERS and sources without well-determined mid-IR and
2-10~keV X-ray luminosities.  This leaves a final sample of 26 local
galaxies with $\nu L_\nu (6 \mu{\rm m}) \simlt 10^{45}\, {\rm erg}\,
{\rm s}^{-1}$.  For the low redshifts of the sample considered here,
the slightly different cosmology adopted by \citet{Gandhi:09} has
negligible effect on our analysis.


The luminosities obtained for the local sample are at $12 \mu$m,
while we require rest-frame luminosities at $6 \mu$m to allow
comparison with the high luminosity samples.  Based on the empirical
AGN template of \citet{Assef:10}, this requires shifting the $12
\mu$m luminosities downwards by 5\%.  Figure~1 presents the X-ray
luminosities plotted against mid-IR luminosity for the local sample.

\subsection{NuSTAR Sample}

Figure~1 also shows seven unbeamed X-ray sources identified from
the {\it Nuclear Spectroscopic Telescope Array}
\citep[\nustar;][]{Harrison:13} serendipitous survey \citep{Alexander:13},
as well as a \nustar-detected source at $z \sim 2$ in the \chandra\
Deep Field-South \citep{DelMoro:14}.  These unobscured or only
somewhat obscured (\ie, Compton-thin; $N_{\rm H} < 1.5 \times
10^{24}\, {\rm cm}^{-2}$) sources all reside within the locus of
broad-lined AGN in the X-ray to mid-IR plane.

Figure~1 also shows several heavily obscured sources studied by
\nustar\ as purple asterisks.  These sources are candidate Compton-thick
AGN ($N_{\rm H} \geq 1.5 \times 10^{24}\, {\rm cm}^{-2}$).  At high
luminosity ($\nu L_\nu (6 \mu{\rm m}) \sim 10^{47}\, {\rm erg}\,
{\rm s}^{-1}$), we show three \wise-selected sources at $z \sim 2$
reported in \citet{Stern:14}; only one of the sources is detected
in the rest-frame 2-10~keV band.  At lower luminosity, we show three
SDSS type-2 AGN discussed in \citet{Lansbury:14}, one of which is
undetected at X-ray energies.  We also include the local luminous
AGN Mrk~34, reported in \citet{Gandhi:14}; this is the \nustar\
source with the lowest X-ray luminosity, $L$(2-10~keV)$= 10^{42}\,
\ergs$.  The \nustar\ heavily obscured sources all reside below
the X-ray to mid-IR relation of unobscured sources, illustrating
how multi-wavelength surveys provide a powerful means of identifying
heavily obscured AGN.

%

\section{The X-Ray to Mid-IR Relation}

\subsection{Literature Relations}

Several groups have previously published the X-ray to mid-IR
correlation for AGN samples.  Using spectral decomposition,
\citet{Lutz:04} separated the nuclear component from the host galaxy
in low resolution spectra of 71 local Seyfert galaxies observed
with the {\it Infrared Space Observatory} ({\it ISO}).  The median
distance of their sources ranges from 50 to 100~Mpc (Seyfert~2 and
Seyfert~1 galaxies, respectively).   Figure~1 presents the correlation
between mid-IR luminosity and absorption-corrected hard (2-10~keV)
X-ray luminosity reported by \citet{Lutz:04} based on their sample.

Rather than using spectral decomposition to separate mid-IR nuclear
emission from contaminating host galaxy emission, \citet{Gandhi:09}
reports on high-resolution imaging obtained with VLT/VISIR to
separate the nuclear and host galaxy contributions to the mid-IR
emission of local AGN.  Based on their cleaned sample of less
obscured local AGN observed with VLT/VISIR, \citet{Gandhi:09} derived
the relation shown in Figure~1.  The consistency with the results
of \citet{Lutz:04} are reassuring given the very different approaches.
However, both samples had few sources with quasar-level X-ray
luminosities, $L$(2-10~keV) $> 10^{44}\, {\rm erg}\, {\rm s}^{-1}$,
and thus were forced to derive a relation with data over a limited
dynamic range, spanning approximately three decades in mid-IR
luminosity.  As clearly seen in Figure~1, these local relations do
a poor job of predicting the mid-IR luminosities of the most luminous
quasars, underpredicting the expected mid-IR luminosity of quasars
with $L$(2-10~keV) $\simgt 10^{45}\, {\rm erg}\, {\rm s}^{-1}$ by
approximately an order of magnitude, or, alternatively, overpredicting
the X-ray luminosities of quasars with $\nu L_\nu(6 \mu{\rm m})
\sim 10^{47}\, {\rm erg}\, {\rm s}^{-1}$ by an order of magnitude.

\citet{Fiore:09} present the X-ray to mid-IR correlation based on
a large sample of X-ray-selected type-1 AGN in the COSMOS and CDF-S
fields using deep {\it Chandra} and {\it Spitzer} observations.  By
considering a sample of sources that reaches into the quasar regime,
the \citet{Fiore:09} relation does a much better job at determining
the X-ray to mid-IR ratio of luminous AGN.  However, there is a
systematic problem with the mid-IR luminosities of the lowest
luminosity AGN being high, most likely due to host galaxy contamination
within the {\it Spitzer} beam.

\citet{Lanzuisi:09} present an X-ray study of a sample of 44 mid-IR
bright ($F_{\rm 24 \mu m} > 1.3$~mJy) objects with extreme
mid-IR-to-optical flux ratios ($F_{\rm 24 \mu m}/F_R > 2000$)
selected from an area of $\sim 6\, {\rm deg}^2$ of the $\sim 50\,
{\rm deg}^2$ \spitzer\ Wide-area InfraRed Extragalactic
\citep[SWIRE;][]{Lonsdale:03} survey with \chandra\ or \xmm\ coverage.
By selection, the sample is biased towards obscured AGN, and their
X-ray to mid-IR correlation, presented in Figure~1, was derived
from the SWIRE sample as well as additional Type~2 quasars from the
literature, where the intrinsic X-ray luminosities were derived
from X-ray spectral fitting.  Perhaps unsurprising given the mid-IR
selection bias to this sample, the \citet{Lanzuisi:09} relation is
slightly offset towards decreased X-ray luminosity (or increased
mid-IR luminosity) relative to the other relations plotted in
Figure~1.

%

\subsection{New Relation}

We derive a new X-ray to mid-IR relation using the data sets from
\S2, which span accurate measurements of these quantities from low
luminosity (\S2.4) to high luminosity (\S2.1).  The sample also
includes several AGN samples of intermediate luminosity.  Doing a
least-square polynomial fit to the data, we derive
$$\log L({\rm 2-10~keV}) = 40.981 + 1.024 x - 0.047 x^2,$$
where $L$(2-10~keV) is in units of $\ergs$ and $x \equiv \log (\nu
L_\nu(6 \mu{\rm m})/10^{41}\, \ergs)$.  If we were to require the
non-local samples to have X-ray luminosities $\log L({\rm 2-10~keV})
> 5 \times 10^{43}\, {\rm erg}\, {\rm s}^{-1}$ so as to avoid sources
with possible host galaxy contributions to their mid-IR emission,
the relation would be essentially unchanged (changes $< 0.2$ dex;
\ie, smaller than the dispersion in the relation shown in Figure~2)
over the range of interest.

Several prior analyses have considered how the optical and X-ray
luminosities of AGN scale over a wide range of AGN power
\citep[\eg,][]{Strateva:05, Steffen:06}.  Such relations are typically
discussed with reference to the parameter $\alpha_{\rm ox}$, defined
by \citet{Tananbaum:79} as the slope of a nominal power law connecting
the rest-frame 2500~\AA\ and 2~keV monochromatic luminosities,
$\alpha_{\rm ox} \equiv 0.3838 \log(L_{\rm 2 keV}/L_{\rm 2500
\mathring{\rm{A}}})$.  For unobscured AGN, $\alpha_{\rm ox}$ compares
the emission power coming from the AGN accretion disk (at rest-frame
2500~\AA) relative to the emission power at rest-frame 2~keV, which
is believed to originate from Compton up-scattering of accretion
disk photons by hot coronal gas of unknown geometry or disk-covering
fraction.  Studies show that as 2500~\AA\ accretion disk
luminosity increases, the luminosity increases less dramatically
at X-ray energies.  The $\alpha_{\rm ox}$ parameter thus probes the
balance between accretion disks and their coronae, and provides a
quantitative constraint on physical models of the structure and
physics of AGN nuclear regions.  The scatter in $\alpha_{\rm ox}$ is
considerable, corresponding to a factor of $\sim 3$ range of X-ray
luminosities observed for sources at a given UV luminosity.  Though
some of this scatter is likely due to non-simultaneous observations
at the two wavelengths, studies show that most of the scatter is
actually intrinsic \citep[\eg][]{Vagnetti:10, Gibson:12}.  \citet{Jin:12}
show that $\alpha_{\rm ox}$ is not correlated with black hole mass, but
is somewhat correlated with Eddington ratio \citep[see also][]{Lusso:10}.
\citet{Steffen:06} reports that there is no significant evolution
of $\alpha_{\rm ox}$ with redshift.


We find a similar behavior here:  as the mid-IR luminosity increases,
the luminosity increases less dramatically at X-ray energies.  AGN
mid-IR emission is generally believed to be primarily due to thermal
radiation from a torus reprocessing accretion disk emission.  Previous
results have shown that the mid-IR to bolometric luminosity of
quasars, $L_{\rm MIR} / L_{\rm bol}$, decreases with increasing
$L_{\rm bol}$, an effect which is generally ascribed to the so-called
receding torus model \citep[\eg][]{Lawrence:91, Simpson:05, Assef:13}:
if the scale height of the torus is independent of the radial size
of the torus, but the inner radius of the torus increases with
increasing AGN luminosity, then the torus effectively covers a
smaller solid angle for more luminous AGN.  Thus, more luminous
quasars are less likely to be seen as obscured, and will also have
relatively diminished thermal IR emission from the obscuring torus.
On the much smaller coronal scale, we also see a trend of decreasing
$L_{\rm X} / L_{\rm bol}$ with increasing $L_{\rm bol}$.  Therefore,
$L_{\rm X} / L_{\rm MIR}$ reflects the competition of physics
occurring on very different size scales, from the sub-parsec coronal
scale, to the much larger obscuring torus scale, though a detailed
investigation and modeling of this, accounting for various selection
effects such as non-simultaneous X-ray and mid-IR imaging, is beyond
the scope of the present work.

In summary, we have derived a new X-ray to mid-IR relation for AGN
which is appropriate for AGN across a large range of luminosity,
from local Seyfert galaxies with $L$(2-10~keV) $\sim 10^{42}\,
\ergs$ out to the most powerful quasars known, with $L$(2-10~keV)
$\sim 10^{46}\, \ergs$.  Previous explorations of this relation
have generally emphasized either just local Seyfert galaxies or
just typical quasars, and no previous analyses have considered the
most luminous quasars.  The result is that extrapolations of the
previous relations differ widely in certain regimes, particularly
at the tip of the luminosity scale.  The revised X-ray to mid-IR
relation for AGN will be beneficial for identifying and studying
highly obscured AGN, which will have their X-ray emission preferentially
suppressed relative to their mid-IR emission.

\acknowledgements

The author gratefully acknowledges communications and input from
close collaborators on the {\it NuSTAR} and {\it WISE} science
teams, particularly David Alexander, Neil Brandt, Peter Eisenhardt,
Poshak Gandhi, Michael Koss, George Lansbury, and Ezequiel Treister.
I am also grateful to the referee, whose suggestions have improved
the paper.  This publication makes use of data products from the
{\it Wide-field Infrared Survey Explorer}, which is a joint project
of the University of California, Los Angeles, and the Jet Propulsion
Laboratory/California Institute of Technology, funded by the National
Aeronautics and Space Administration.  The author also acknowledges
support from NASA through ADAP award 12-ADAP12-0109.

\smallskip
{\it Facilities:} \facility{Chandra}, \facility{Spitzer},
\facility{WISE}, \facility{XMM-Newton}

\smallskip
\copyright 2015 California Institute of Technology.  Government sponsorship acknowledged.

\clearpage
\end{document}